\documentclass[aps,prl,superscriptaddress,twocolumn,showpacs,preprintnumbers,amsmath,amssymb]{revtex4}

\usepackage[dvips]{graphicx,color}
\usepackage{dcolumn}
\usepackage{bm}
\usepackage{ulem}
\usepackage{soul}
\begin{document}

\title{Giant Spin Hall Effect Induced by Skew Scattering from Bismuth Impurities inside Thin Film CuBi Alloys}

\author{Y. Niimi}
\email{niimi@issp.u-tokyo.ac.jp}
\affiliation{Institute for Solid State Physics, University of Tokyo, 5-1-5 Kashiwa-no-ha, Kashiwa, Chiba 277-8581, Japan}
\author{Y. Kawanishi}
\affiliation{Institute for Solid State Physics, University of Tokyo, 5-1-5 Kashiwa-no-ha, Kashiwa, Chiba 277-8581, Japan}
\author{D. H. Wei}
\affiliation{Institute for Solid State Physics, University of Tokyo, 5-1-5 Kashiwa-no-ha, Kashiwa, Chiba 277-8581, Japan}
\author{C. Deranlot}
\affiliation{Unit\'{e} Mixte de Physique CNRS/Thales, 91767 Palaiseau France associ\'{e}e \`{a} l'Universit\'{e} de Paris-Sud, 91405 Orsay, France}
\author{H. X. Yang}
\affiliation{SPINTEC, UMR CEA/CNRS/UJF-Grenoble 1/G-INP, INAC, 38054 Grenoble, France}
\author{M. Chshiev}
\affiliation{SPINTEC, UMR CEA/CNRS/UJF-Grenoble 1/G-INP, INAC, 38054 Grenoble, France}
\author{T. Valet}
\affiliation{In Silicio SAS, 730 rue Ren\'{e} Descartes, 13857 Aix en Provence Cedex 3, France}
\author{A. Fert}
\affiliation{Unit\'{e} Mixte de Physique CNRS/Thales, 91767 Palaiseau France associ\'{e}e \`{a} l'Universit\'{e} de Paris-Sud, 91405 Orsay, France}
\author{Y. Otani}
\affiliation{Institute for Solid State Physics, University of Tokyo, 5-1-5 Kashiwa-no-ha, Kashiwa, Chiba 277-8581, Japan}
\affiliation{RIKEN-ASI, 2-1 Hirosawa, Wako, Saitama 351-0198, Japan}

\date{October 9, 2012}

\begin{abstract}
We demonstrate that a giant spin Hall effect (SHE) can be induced 
by introducing a small amount of Bi impurities in Cu. Our analysis based 
on a new 3-dimensional finite element treatment of spin transport shows 
that the sign of the SHE induced by the Bi impurities is negative and 
its spin Hall (SH) angle amounts to $-0.24$. Such a negative large SH angle 
in CuBi alloys can be explained by applying the resonant scattering model 
proposed by Fert and Levy [Phys. Rev. Lett. {\bf 106}, 157208 (2011)] 
to 6$p$ impurities.
\end{abstract}


\maketitle

Spintronic devices manipulating pure spin currents, 
flows of spin angular momentum without charge current, 
should play an important role in low energy consumption electronics 
of next generation. This explains the current interest for 
the spin Hall effect (SHE) which provides a purely electrical way 
to create spin currents without ferromagnets and magnetic fields.
The SHE, originally predicted by Dyakonov and Perel~\cite{dp_1971}, 
can be described as an accumulation of spins generated by an electric current 
on the edges of a nonmagnetic conductor~\cite{kato_science_2004}. 
Its interest in spintronics comes from its application 
to convert charge into spin currents (or vice-versa). 
The spin Hall (SH) angle, 
characteristic of the conversion yield between charge and spin, reaches a 
few \% in heavy metals with strong spin-orbit (SO) interactions such as 
Pt~\cite{saitoh_apl_2006,kimura_prl_2007,vila_prl_2007,hoffmann_prl_2010,liu_prl_2011,morota_prb_2011}.

\begin{figure}
\begin{center}
\includegraphics[width=8.5cm]{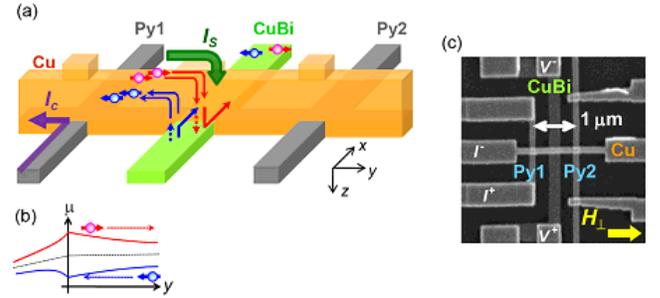}
\caption{(Color online) (a) Schematic of the ISHE measurement. The ISHE in CuBi deflects spin-up and down electrons $|e|$ ($e$ is the charge of the electron) denoted by spheres with arrows to the same side. Other arrows indicate the electron motion direction. (b) Electrochemical potential distributions for spin-up and down electrons in Cu. Spin-up and down electrons diffuse in opposite directions, which characterizes a pure spin current (no charge current). (c) Scanning electron microscopy image of our spin Hall device. The yellow arrow in (c) represents the positive direction of the magnetic field in the case of ISHE measurement.} \label{fig1}
\end{center}
\end{figure}

In addition to the intrinsic SHE of 
pure metals~\cite{morota_prb_2011,tanaka_prb_2008}, 
the skew scattering~\cite{skew} and the scattering with 
side-jump~\cite{side_jump} on impurities 
with strong SO interactions can also give rise to the SHE. 
Such extrinsic mechanisms have already been studied
in the SHE induced by nonmagnetic impurities in Cu~\cite{fert_jmmm_1981} 
and in the anomalous Hall effect of ferromagnetic alloys~\cite{fert_prl_1972}. 
One of the big advantages for the extrinsic SHE 
is that it allows controlling the SH angle by changing the combination 
of host and impurity metals as well as by tuning the impurity concentration. 
According to recent theoretical 
predictions~\cite{gradhand_prl_2010,fert_prl_2011}, 
some combinations of noble metals and impurities could induce very 
large SH angles, for example in Cu or Ag doped with Bi. 
In this work we obtain a large SHE signal in Cu doped 
with a small amount ($\leq 0.5$\%) of Bi. 
From our analysis based on a 3-dimensional (3D) 
finite element treatment of the spin transport 
equations~\cite{valet_prb_1993}, the SH angle can be 
estimated to be $-0.24$ at $T = 10$~K, which is an order of magnitude 
larger than that in pure metals such as Pt, Pd and almost twice larger than 
that announced recently for the $\beta$ phase of Ta~\cite{liu_science_2012}.

Figure~1(a) shows the principle of the ISHE using the spin absorption 
method~\cite{kimura_prl_2007,vila_prl_2007,morota_prb_2011,niimi_prl_2011,supplemental_material}. 
When the electric current flows from Py1 to the left side of the Cu wire, 
the resulting spin accumulation induces a pure spin current 
(no net charge current, i.e. $I_{\rm C} = 0$ or 
$I_{\uparrow} = -I_{\downarrow}$ for opposite flows of spin-up and 
down electrons) on the right side of the Cu wire [see Fig.~1(b)]. 
As discussed later on, most of the pure spin current is absorbed vertically 
into the middle wire below Cu. The opposite spin-up and down electrons 
composing the absorbed pure spin current are deflected to 
the same direction (along the $x$-axis) by the ISHE in CuBi, 
and an ISHE voltage $V_{\rm ISHE}$ is generated to prevent the flow of 
a charge current along the $x$ direction in CuBi. 
By inverting the probe configuration (i.e. $I^{+} \Leftrightarrow V^{+}$, 
$I^{-} \Leftrightarrow V^{-}$ in Fig. 1(c)), one can also measure the direct
SHE (DSHE)~\cite{kimura_prl_2007,vila_prl_2007,niimi_prl_2011}; 
with an electric current in the CuBi wire, the spin accumulation induced at 
the interface between Cu and CuBi can be detected from the nonlocal 
voltage between Py1 and Cu.

\begin{figure}
\begin{center}
\includegraphics[width=6cm]{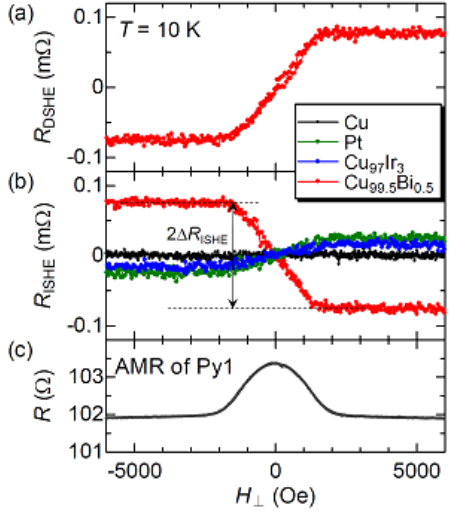}
\caption{(Color online) (a) DSHE and (b) ISHE signals of Cu$_{99.5}$Bi$_{0.5}$ measured at $T = 10$~K. For comparison, the ISHE resistances of a 20~nm thick pure Cu ($\rho = 6.3$~$\mu\Omega\cdot$cm), Pt (10~$\mu\Omega\cdot$cm), and Cu$_{\rm 97}$Ir$_{3}$ (14~$\mu\Omega\cdot$cm) are also added in (b). (c) A typical AMR signal of Py1 showing the saturation of the magnetization above 2000~Oe for $H_{\perp}$ along the $y$ direction [see Fig.~1(a)].} \label{fig2}
\end{center}
\end{figure}

The ISHE resistance $R_{\rm ISHE}$ ($\equiv V_{\rm ISHE}/I_{\rm C}$) 
as well as the DSHE resistance $R_{\rm DSHE}$ for Cu$_{99.5}$Bi$_{0.5}$ 
are plotted in Fig~2, with also reference signals for pure Cu, Pt, 
and Cu$_{97}$Ir$_{3}$. The SHE of Cu is negligibly small but, 
once only a small amount of Bi is added in Cu, 
the alloy shows a quite large SHE signal. $R_{\rm ISHE}$ linearly 
increases with increasing the magnetic field and it is saturated 
above 2000~Oe which is the saturation field of the magnetization, 
as can be seen in the anisotropic magnetoresistance (AMR) curve of Fig.~2(c). 
The amplitude of the SHE resistance is exactly the same for 
both the DSHE and ISHE, in agreement with the Onsager reciprocal 
relation~\cite{kimura_prl_2007,vila_prl_2007,niimi_prl_2011}. 
We note the following two points: (I) the sign of $R_{\rm ISHE}$ 
for CuBi is opposite to that for 
Pt~\cite{kimura_prl_2007,vila_prl_2007,morota_prb_2011} and 
Cu$_{97}$Ir$_{3}$~\cite{niimi_prl_2011}, and 
(II) the amplitude ($\Delta R_{\rm ISHE}$) of the ISHE resistance 
for Cu$_{99.5}$Bi$_{0.5}$ is several times larger than that of Pt and 
Cu$_{97}$Ir$_{3}$ although their residual resistivities are 
almost the same ($\rho \sim 10~\mu\Omega\cdot$cm), indicating that 
the SH angle in CuBi is much larger than in Pt and CuIr.

\begin{figure}
\begin{center}
\includegraphics[width=5.5cm]{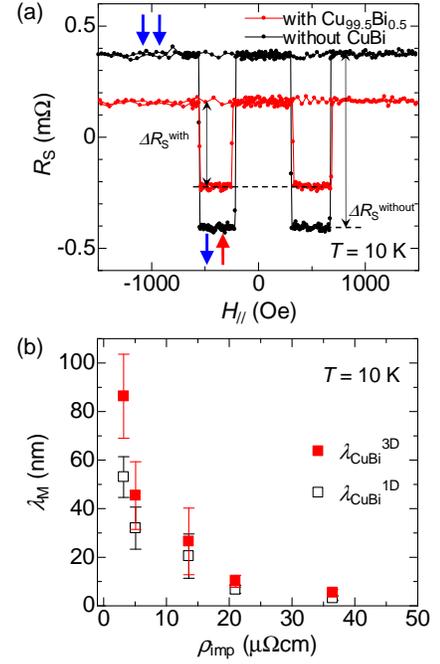}
\caption{(Color online) (a) NLSV signals measured at $T = 10$~K with a Cu$_{99.5}$Bi$_{0.5}$ middle wire (red) and without CuBi wire (black). Arrows represent the magnetization directions of Py1 (left) and Py2 (right). Note that for the NLSV measurement the magnetic field is aligned along the easy axis ($x$ direction) of the Py wires. (b) Spin diffusion length $\lambda_{\rm M}$ of CuBi alloys at 10~K as a function of $\rho_{\rm imp}$. Closed and open symbols correspond respectively to the 3D and 1D analyses.} \label{fig3}
\end{center}
\end{figure}

For a quantitative analysis, one has to know how much of 
the pure spin current $I_{\rm S}$ generated from Py1 is absorbed 
into the CuBi middle wire. As mentioned above, most of the spin current 
flowing in Cu is injected into the CuBi wire because of its strong 
SO interaction, but the rest of the spin current also flows toward 
the second Py wire (Py2). 
By measuring nonlocal spin valve (NLSV) signals with and without 
the middle wire, one can obtain the spin absorption rate into the CuBi 
wire~\cite{vila_prl_2007,morota_prb_2011,niimi_prl_2011}. 
Figure 3(a) shows the NLSV signals with and without 
the Cu$_{99.5}$Bi$_{0.5}$ wire. With the CuBi wire, 
the spin accumulation signal $\Delta R_{\rm S}^{\rm with}$, 
defined by the spin accumulation voltage ($\Delta V_{\rm S}^{\rm with}$) 
divided by $I_{\rm C}$, is reduced to $\sim50$\% compared to that without 
the CuBi wire ($\Delta R_{\rm S}^{\rm without}$). 
The difference between $\Delta R_{\rm S}^{\rm without}$ and 
$\Delta R_{\rm S}^{\rm with}$ depends strongly on the spin diffusion length 
of CuBi, $\lambda_{\rm M}$, and can be used to extract $\lambda_{\rm M}$, 
what we have done in two types of model: 
(A) the standard 1-dimensional (1D) model of diffusive 
spin transport developed by Takahashi and 
Maekawa~\cite{takahashi_prb_2003} and already used 
in publications by some of us~\cite{morota_prb_2011,niimi_prl_2011}, and 
(B) a 3D finite element treatment of diffusive spin transport based on 
an extension of the Valet-Fert formalism to non-collinear spins, 
SHE and 3D distributions of spin and charge currents 
(SpinFlow 3D~\cite{valet_prb_1993}). 
More details on the two models are presented 
in supplemental material~\cite{supplemental_material} 
with a list of the parameters which we use in the 1D and 3D calculations. 
From the data on $\Delta R_{\rm S}^{\rm without}$ and 
$\Delta R_{\rm S}^{\rm with}$ we have derived $\lambda_{\rm M}$ 
in the 1D and 3D models and we present the results in Fig.~3(b). 
In both models $\lambda_{\rm M}$ rapidly decays with increasing 
the Bi concentration, which expresses the contribution of the Bi impurities 
to SO-induced spin relaxation. The 3D model gives a longer spin diffusion 
length for $\lambda_{\rm M} > t_{\rm M}$ (thickness of the CuBi wire). 

Once $\lambda_{\rm M}$ is known, 
we have used its respective values from the 1D and 3D calculations 
to derive the SH resistivity $\rho_{\rm SHE}$ in both approaches. 
In the 1D approach, we used the following standard equation 
obtained by combining Eqs.~(2) and (3) of Ref.~\cite{niimi_prl_2011}:
\begin{widetext}
\begin{eqnarray}
\rho_{\rm SHE} \simeq \Delta R_{\rm ISHE} 
\frac{w_{\rm M}}{x} \frac{t_{\rm M}}{\lambda_{\rm M}} 
\frac{1-\exp\left(-2t_{\rm M}/\lambda_{\rm M} \right)}{\left\{ 1-\exp\left(-t_{\rm M}/\lambda_{\rm M}\right) \right\}^{2}} \frac{R_{\rm N} \left\{ \cosh \left( L/\lambda_{\rm N} \right) - 1 \right\} + 2 R_{\rm F} \left\{ \exp\left( L/\lambda_{\rm N} \right) - 1 \right\} +2 R_{\rm M} \sinh \left( L/\lambda_{\rm N} \right)} {2 p_{\rm F} R_{\rm F} \sinh \left( L/2\lambda_{\rm N} \right)}, \label{eq3}
\end{eqnarray}
\end{widetext}
where $L$, $w_{\rm M}$, $\lambda_{\rm N}$, and $p_{\rm F}$ are respectively 
the distance between Py1 and Py2 (fixed to 1~$\mu$m), 
the width of the CuBi wire, the spin diffusion length of Cu, 
and the spin polarization of Py, 
while the spin resistances $R_{\rm X}$ ($X =$ N, F and M) are defined 
in Ref.~\cite{supplemental_material}. 
The determination of 
the shunting coefficient $x$ has been the object of a recent 
debate~\cite{liu_arxiv_2011} and this is one of the reasons for 
which we have introduced a 3D modeling to take automatically into account 
the shunting. 
As we will see later on, however, 
the larger deviation of the results of the 1D model from those of the 3D one 
comes less from the shunting effects than from the spreading of 
the spin accumulation over the sides of the contacts 
in the SHE material.

\begin{figure*}
\begin{center}
\includegraphics[width=13.5cm]{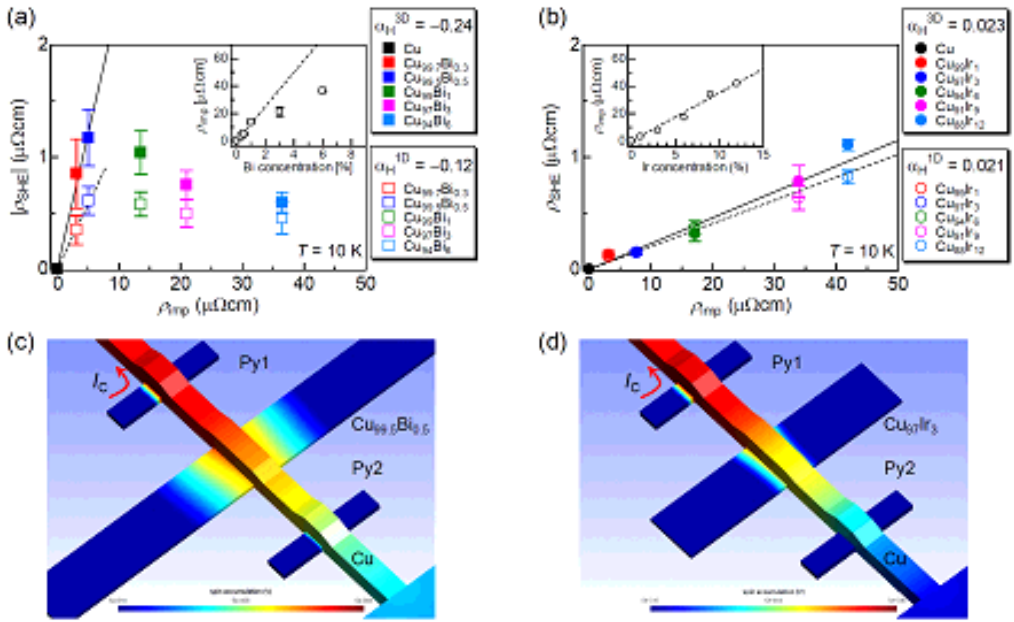}
\caption{(Color online) (a) SH resistivity of CuBi alloys as a function of $\rho_{\rm imp}$. Closed and open symbols correspond respectively to the 3D and 1D analyses. The SH angles $\alpha_{\rm H}^{\rm 3D}$ and $\alpha_{\rm H}^{\rm 1D}$ correspond to the slopes of solid and broken lines, respectively. The inset shows the resistivity induced by Bi impurities $\rho_{\rm imp}$ as a function of Bi concentration. (b) For comparison, the SH resistivity of CuIr alloys is plotted as in (a), but the experimental data are derived from Fig. 3 in Ref.~\cite{niimi_prl_2011}. The linear variation of the SH resistivity with $\rho_{\rm imp}$ is not limited by the solubility of Ir up to concentrations as large as 12\%. (c), (d) 3D mappings of the spin accumulation voltage for the (c) Cu$_{99.5}$Bi$_{0.5}$ and (d) Cu$_{97}$Ir$_{3}$ SH devices calculated with SpinFlow 3D~\cite{valet_prb_1993}.} \label{fig4}
\end{center}
\end{figure*}

In Fig.~4(a) the SH resistivities of the CuBi alloys derived 
by our 1D and 3D calculations are plotted as a function of the resistivity 
induced by the Bi impurities, 
$\rho_{\rm imp} \equiv \rho_{\rm CuBi} - \rho_{\rm Cu}$. 
For both calculations the linear variation of the SH resistivity 
characteristic of skew scattering by dilute impurities is observed 
only at the lowest concentrations. The large deviation from linearity 
from $c = 1$\% and above is consistent with a similar deviation 
from linearity in the variation of the CuBi resistivity 
with the Bi concentration [see the inset of Fig.~4(a)] and 
simply reflects the departure from the dilute impurity regime. 
We present in Ref.~\cite{supplemental_material} 
scanning tunneling electron microscopy (STEM) and 
energy dispersive X-ray (EDX) data showing the inhomogeneous distribution 
on Bi in the concentrated alloys~\cite{cubi_phase_diagram}. 
Note that the dilute impurity regime is more extended 
in the CuIr case~\cite{niimi_prl_2011}, where both $\rho_{\rm SHE}$ and 
$\rho_{\rm imp}$ follow linear variations up to Ir concentration 12\% 
[see Fig.~4(b)].

From now we consider only the dilute impurity regime characterized 
by linear variations of the resistivity and SH resistivity 
with the concentration. 
Considering the slope $\rho_{\rm SHE}/\rho_{\rm imp}$ in this regime, 
that is the SH angle $\alpha_{\rm H}$ characteristic of the skew scattering 
by Bi impurities~\cite{niimi_prl_2011}, we remark that the 3D calculation
gives a larger SH angle, $\alpha_{\rm H}^{\rm 3D} = -(0.24 \pm 0.09)$
compared to $\alpha_{\rm H}^{\rm 1D} = -(0.12 \pm 0.04)$. 
The main reason for this underestimation of the SH angle in the 1D approach is 
obvious from Fig.~4(c) where one sees the 3D spreading of 
the spin accumulation on both side edges at the CuBi/Cu junction. 
This spreading is important when $\lambda_{\rm M}$ is longer than $t_{\rm M}$. 
Calculations with the 1D model, in which a pure spin current is injected 
\textit{only} vertically into the CuBi wire, cannot take into account 
the spin escape by lateral spreading and thus underestimate the SH angle. 
The 3D modeling allows to calculate precisely all the contributions 
coming from local distributions of current and shunting effects 
around the contacts.
The difference between the two models is much smaller 
for the CuIr alloys where $\lambda_{\rm M}$ is in general shorter 
than $t_{\rm M}$ and the spin accumulation spreading is less important 
[see Figs.~4(b) and 4(d)]. 
The situation is similar for our Pt wire 
($\lambda_{\rm Pt} \sim 10$~nm $< t_{\rm Pt} = 20$~nm, 
$\alpha_{\rm H}^{\rm 1D} = 0.021$ and $\alpha_{\rm H}^{\rm 3D} = 0.024$). 
These results are in contrast to the propositions made by 
Liu \textit{et al.}~\cite{liu_arxiv_2011}; 
taking into account the 3D effects 
(shunting, spin accumulation spreading, etc) does not lead to a 
shorter $\lambda_{\rm M}$. 
On the contrary $\lambda_{\rm M}$ is slightly longer in our 3D treatment, 
and thus the 1D model underestimates the SH angle 
(in the present case by a factor of 2), not when $\lambda_{\rm M}$ 
is shorter than $t_{\rm M}$ but when it is longer. 
Thus, the important result is the giant SH angle, 
$-0.24$, induced by skew scattering on Bi impurities in Cu. 
This SH angle is the largest value as far as we know, 
definitely above the range $0.12-0.15$ recently found in 
Au~\cite{bogu_prl_2010} and Ta~\cite{liu_science_2012}. 
Note that $\alpha_{\rm H} = -0.24$, obtained by dividing $\rho_{\rm SHE}$ 
by $\rho_{\rm imp}$, is the characteristic SH angle of the skew scattering 
on Bi. When $\rho_{\rm SHE}$ is divided by 
$\rho_{\rm CuBi} (= \rho_{\rm Cu} + \rho_{\rm imp}$), 
the global SH angle of the alloy 
becomes $-0.11$. The value $\alpha_{\rm H} = -0.24$ would be obtained 
by using the copper of smaller residual resistivity 
(less defect and thicker film).

The large SH angle predicted by a recent \textit{ab}-initio calculation 
for CuBi alloys~\cite{gradhand_prl_2010} is consistent with 
our experiment but the signs are opposite. To clear up the issue, 
we have tried another theoretical approach adapting the phase shift model 
worked out by Fert and Levy~\cite{fert_prl_2011} for 5$d$ impurities to 
the case of Bi. We start with an \textit{ab}-initio calculation 
(using Quantum-ESPRESSO~\cite{quantum_espresso}) of 
the numbers of electrons at $j = 1/2$ and $j = 3/2$ 
$p$-states in the conduction band on the Bi site 
(see Ref.~\cite{supplemental_material} for more details). 
By comparing with the numbers of $p$-conduction electrons on a Cu site, 
we find the numbers of electrons attracted on a Bi site and 
derive the $j = 1/2$ and $j = 3/2$ scattering phase shifts 
($\eta_{1/2}$ and $\eta_{3/2}$) from Friedel's sum rule. 
The difference in these phase shifts comes from the SO interaction 
and induces the SHE.
The phase shift $\eta_{0}$ of the $l = 0$ channel is also found 
from the \textit{ab}-initio calculations. 
A straightforward extension of the formalism developed in 
Ref.~\cite{fert_prl_2011} for $d$ to $p$ states leads to the 
following expression of the SH angle:
\begin{widetext}
\begin{eqnarray}
\alpha_{\rm H}= \frac{-2\sin\eta_{0}\left\{ \sin\eta_{1/2}\sin\left(\eta_{1/2}-\eta_{0} \right) - \sin\eta_{3/2}\sin\left(\eta_{3/2} -\eta_{0} \right)  \right\}}{3\left(\sin^{2}\eta_{0}+\sin^{2}\eta_{1/2}+2\sin^{2}\eta_{3/2} \right)}. \label{eq2}
\end{eqnarray}
\end{widetext}
The final result for the SH angle of CuBi alloy is $-0.046$. 
The negative sign is in agreement with our experimental results and 
the SH angle is large but not as large as in the experiments. 
However, it can be further increased, for example, 
by taking into account electron correlation effects which may significantly 
enhance the SO splitting as recently proposed by 
Gu \textit{et al}~\cite{bogu_prl_2010}. Indeed, we have checked that 
an increase of the splitting between the $\eta_{1/2}$ and $\eta_{3/2}$ 
phase shifts by a factor of about 3 keeps the right sign but would bring 
the SH angle beyond $-0.26$. 

Finally one wonders if interfacial effects could also contribute 
to our experimental results. The possible competition between the SHE 
and the interfacial Rashba effects has indeed been debated 
recently~\cite{miron_nature_2011,liu_prl_2012} 
for the interpretation of experiments 
with $2\sim 3$~nm thick Pt layers. For our thicker CuBi films (20~nm), 
the SHE contribution can be expected to be largely predominant. 
Also some enhancement of the SHE 
at interfaces~\cite{Hou_apl_2012} can be neglected in a 20~nm thick 
metallic film in which the interface states cannot exceed a couple 
of atomic layers.

In conclusion, we find that a small amount of Bi impurities ($\leq 0.5$\%) 
in Cu induces a large SHE. We have analyzed our data using both the classical 
1D model of previous SHE studies and a 3D finite element treatment of 
spin transport. A large SH angle is derived from both models. 
It is definitely larger in the 3D model ($-0.24$). Such a difference 
between 1D and 3D models is not surprising 
since the 3D model can treat more accurately the unavoidable approximations 
of the 1D model. 
Harnessing such a giant SHE to produce or detect spin currents 
will be probably more and more used in novel generations of 
spintronic devices not necessarily based on magnetic materials.

We acknowledge helpful discussions with I. Mertig, M. Gradhand, 
S. Maekawa, T. Ziman, B. Gu, and A. Smogunov. 
We would like to thank Y. Iye and S. Katsumoto 
for the use of the lithography facilities and also P. M. Levy 
for the derivation of Eq.~(\ref{eq2}). This work was supported by KAKENHI, 
the Agence Nationale de la Recherche, and 
a Grant-in-Aid for Scientific Research in Priority Area from MEXT.

\end{document}